# The dM4.5e star G124-62 and its binary L dwarf companion DENIS-P J1441-0945⋆

## Common proper motion, distance, age, and masses

Andreas Seifahrt[1,2], Eike Guenther[3], and Ralph Neuhäuser[1]

[1] Astrophysikalisches Institut, Universität Jena, Schillergässchen 2-3, 07745 Jena, Germany
[2] European Southern Observatory, Karl-Schwarzschild-Str. 2, 85748 Garching
[3] Thüringer Landessternwarte Tautenburg, Sternwarte 5, 07778 Tautenburg, Germany

**Abstract.** We present astrometric and photometric measurements of the field binary L1 dwarf DENIS-P J1441-0945. Analysis of archival HST images and photometric parallax measurements give a distance of 34±7 pc and a proper motion close to that of the nearby high proper motion star G124-62. Comparison of SuperCOSMOS and 2MASS images confirms that these objects form a common proper motion pair, while spectroscopy of G124-62 shows it to be a dM4.5e star. The kinematics show that this system is a member of the Hyades supercluster. The resulting age constraints for the system are 500–800 Myr and the mass of each component of DENIS-P J1441-0945 is $0.072^{+0.010}_{-0.018}$ M$_\odot$.

## 1. Introduction

Once formed, brown dwarfs cool over time and never reach the main sequence (Burrows, Hubbard, Lunine, & Liebert, 2001). This leads to a degeneracy in the determination of age and mass for models, which can only be lifted for brown dwarf binaries in close orbits and for brown dwarf companions to other stars. This paper describes new data on the binary L1 dwarf DENIS-P J1441-0945 (Martín et al., 1999; Bouy et al., 2003) which provide measurements of these quantities.

The lack of Li I ($\lambda$ 6708 Å) absorption sets limits of $M \gtrsim 0.05 M_\odot$ and age $\gtrsim 0.5$ Gyr (Martín et al., 1999) for DENIS-P J1441-0945, since only the most massive brown dwarfs burn lithium up to a certain age. Bouy et al. (2003) find that the object is an equal-luminosity binary with a separation of 375 ± 2.8 mas at a photometric distance of 29 pc. The resulting linear separation implies an orbital period of $\gtrsim 100$ yr based on this distance and an assumed mass of 0.1 $M_\odot$ for each component, so that it will take several decades of observation before the mass can be measured directly.

Because it is 44 arcsec northeast of the high proper motion star G124-62, DENIS-P J1441-0945 is a possible companion to this star (see J. D. Kirkpatrick's web page[1]). In this paper we use astrometry from the HST WFPC2 archival images, 2MASS, and SuperCOSMOS to confirm that these objects form a common proper motion pair and to measure the parallax for the system. We also describe a high dispersion spectrum of G124-62, from which we determine its spectral type, measure its radial velocity, and search for chromospheric activity. Together these data show that the system is a member of the Hyades supercluster and allow us to derive a good estimate of the age, and hence the masses, of the components of DENIS-P J1441-0945. The astrometric data are discussed in Sect. 2 and the spectroscopic data in Sect. 3.1. In Sect. 3.2, we derive constraints on the age of the system, which are used in Sect. 4 to derive the masses of the components of DENIS-P J1441-0945, showing that these objects lie close to the stellar-substellar boundary. The conclusions are presented in Sect. 5.

---

[1] `http://spider.ipac.caltech.edu/staff/davy/ARCHIVE/`

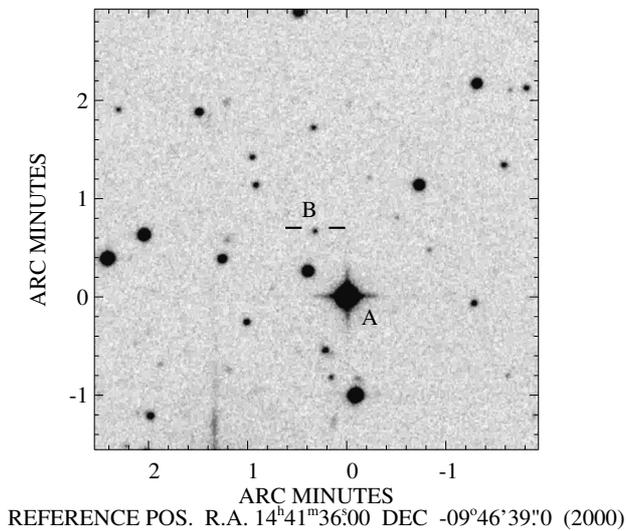

**Fig. 1.** SSS images showing the field of (A) G124-62 and (B) DENIS-P J1441-0945. North is up and east to the left.

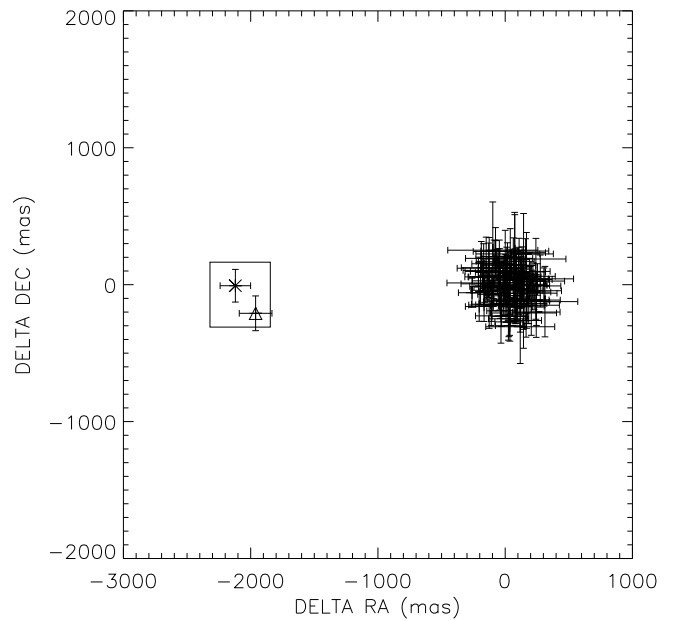

**Fig. 2.** Measured position differences between SSS image (April, 27 1989) and 2MASS catalogue (March, 14 1999). G124-62 is marked as a cross and DENIS-P J1441-0945 as a triangle. The expected displacement based on the NLTT proper motions is shown by the box, $1\sigma$ errors are plotted for all values.

## 2. Data on DENIS-P J1441-0945

### 2.1. DENIS-P J1441-0945 on the SuperCOSMOS Sky Survey

The SuperCOSMOS Sky Survey (hereafter SSS) includes plates from the UK Schmidt Telescope Southern Survey (hereafter UKST) in I band scanned by SuperCOSMOS at the University of Edinburgh. These plates have a limiting magnitude of 19.5 and a resolution of 0.7 arcsec/pixel. DENIS-P J1441-0945 has an I band magnitude of 17.32 (Martín et al., 1999) and is easily detectable on the corresponding SSS image (see Fig. 1).

G124-62 is listed in the NLTT catalogue with a proper motion of 211 ± 24 mas/yr ($\theta$ = 268 ± 7 deg) [2] and in the Lowell Proper Motion catalogue with 270 ± 30 mas/yr ($\theta$ = 264 ± 7 deg), respectively, i.e. almost directly east-west. The epoch difference between the SSS image and the 2MASS catalogue is nearly 10 years resulting in an expected motion on the plane of sky of ~ 2″.

In Figure 2 we show the displacements between the positions in these catalogues. The overlapping error bars of DENIS-P J1441-0945 and G124-62 demonstrate that these objects share the same proper motion within the 1 $\sigma$ errors. The proper motion of both objects differs by more than 15$\sigma$ from the background objects, hence, G124-62 and DENIS-P J1441-0945 form a common proper motion pair. We refer to DENIS-P J1441-0945 hereafter as G124-62B and the primary as G124-62A. Our measured proper motion is 206 ± 13 mas/yr ($\theta$ = 267 ± 4 deg) for this system. For a detailed discussion of the techniques of measurement see Seifahrt, Neuhäuser, & Mugrauer (2004).

### 2.2. DENIS-P J1441-0945 on the HST/WFPC2

Archived data from *HST*/WFPC2 for G124-62B are available from the programs GO8720 (cycle 9, PI W. Brandner), GO9157 (cycle 10, PI E. M. Martín), and GO9345 (cycle 11, PI E. M. Martín). Five observing runs span a range in observation time from January 16, 2001 to January 01, 2003.

G124-62A is saturated in all of these images. G124-62B is easily detected and is resolved into two components (see Fig. 3), hereafter called G124-62Ba and Bb. The position change of the system due to proper motion can be seen even in the short time interval of the WFPC2 observations. Furthermore, we found that the position accuracy is sufficient to measure the parallax of G124-62B. We re-aligned the images by overlaying the positions of the background stars in each image. Since there are not enough objects in the small Planetary Camera field we had to use the Wide Field Camera images, although its pixel scale is larger (100 mas/pixel versus 45.5 mas/pixel of the PC field), resulting in errors for the position changes in both RA and DEC of ~ 13 mas (1$\sigma$). Figure 4 shows the position differences in RA[3] relative to the first epoch observation (Mai 22, 2001). Fitting the data by applying a least-square method results in a proper motion of $\mu_\alpha$ = −197 ± 14 mas/yr and a parallax of $\pi$ = 30 ± 17 mas, giving a distance of $d = 33^{+44}_{-12}$ pc, consistent within the errors with the photometric distance of 29.2 pc measured by Bouy et al. (2003). Since the parallax-measurements are linked to the proper motion measurements, independent determination of the proper motions in

---

[2] The error given is the typical 1 $\sigma$ error in proper motion for the NLTT catalogue according to Gould & Salim (2003).

[3] Since the object is near the ecliptic the parallax motion in DEC is negligible.

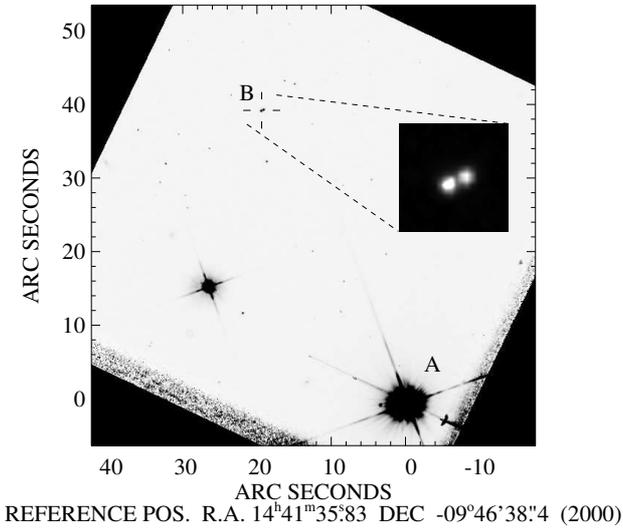

**Fig. 3.** Archived WFPC2 image of (A) G124-62 and (B) DENIS-P J1441-0945 from January 01, 2003. All other objects are background objects. North is up and east to the left. Subset: DENIS-P J1441-0945 is clearly resolved into two components. Image obtained by PI E. M. Martín.

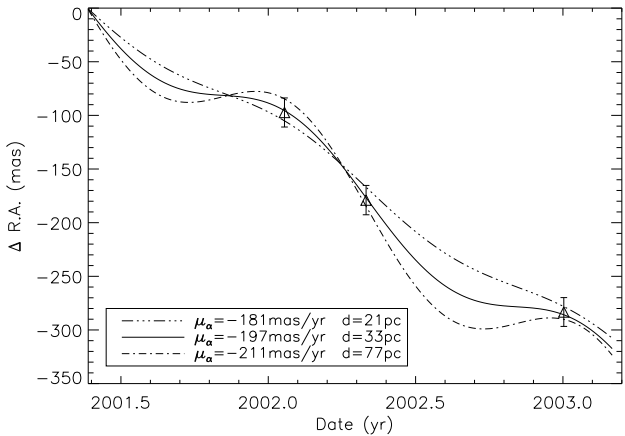

**Fig. 4.** Parallax measurement. See text for discussion.

Sect. 2.1 further constrains the parallax. In this case the proper motion gives us an additional lower limit for the parallax, resulting in the best value for the distance of $d = 33^{+24}_{-12}$ pc.

Unfortunately no error is provided for the photometric distance given by Bouy et al. (2003). Therefore we redetermined the spectrophotometric distance of G124-62B in order to derive a second and independent value for the distance of the system.

Martín et al. (1999) determined the spectral type of G124-62B to L1 ± 0.5 from spectroscopy. This can be adopted for both G124-62Ba and Bb, since both objects show only small differences in brightness in two filters (Bouy et al., 2003). A relation between absolute magnitude and spectral type for late-M and L dwarfs, given by Dahn et al. (2002)

$$M_J = 8.38 + 0.341 \times SpT , \quad \sigma = 0.25 \quad (1)$$

(where SpT is 7 for M7, 18 for L8) implies an absolute magnitude in the $J$ band of $M_J = 12.13 \pm 0.42$ mag for the given spectral type. The 2MASS magnitudes in $J$ and $K$ band of $J = 14.02 \pm 0.026$ mag and $K = 12.661 \pm 0.029$ mag have to be corrected for the duplicity of the object. Assuming equal brightness for both components, their magnitudes were overestimated by 0.753 mag. Using the corrected magnitudes we can derive a spectrophotometric distance to G124-62Ba and Bb of $d = 34 \pm 7$ pc. This agrees well with the distance derived from our parallax measurements and the distance given in Bouy et al. (2003).

Using this distance we find an absolute magnitude in the $K$ band of $M_K = 10.77 \pm 0.48$ mag for each component. Alternatively, we can use the relationship given in Reid & Cruz (2002) to derive $M_K$:

$$M_K = 7.593 + 2.25 \times (J - K_S) , \quad \sigma = 0.36 \quad (2)$$

Using this relation we find only a slightly different value of $M_K = 10.65 \pm 0.48$ mag. Both values are consistent within the error margins. Concluding on the distance of the system by photometric measurements would therefore result in the same value and similar uncertainties. The distances determined spectroscopically and by measuring the parallax agree well with each other.

## 3. Data on G124-62A

Only very little information is available on G124-62A. Apart from the astrometric data from the NLTT catalogue and the Lowell Proper Motion catalogue, photometric data is also given in the Tycho2 catalogue and in the 2MASS catalogue. The spectral type K, listed in SIMBAD, originated from the colour class assignment in the NLTT catalogue and is therefore uncertain. The All Sky Automated Survey (hereafter ASAS; Pojmanski 2002) finds no $V$ band variability greater than 0.1 mag between March 29, 2001 and July 3, 2004.

### 3.1. Spectral type

To further investigate the nature of G124-62A we obtained a spectrum of this object with FEROS on the ESO 2.2m Telescope. The resulting spectrum has a resolution of ∼48.000 and shows absorption lines of CaI, FeI, VI, and TiO, as well as a strongly broadened Na D doublet. Hydrogen, Helium, CaII H & K, Na D, and CaII ($\lambda\lambda$ 8498, 8542 and 8662Å) lines are in emission.

According to the classification scheme of Kirkpatrick, Henry, & McCarthy (1991), the spectrum of G124-62A is an early-to-mid M dwarf. Additional parameters, like the ratio of H$\alpha$ to H$\beta$ emission (Gizis, Reid, & Hawley, 2002), further constrain the spectral type, which is then best described as dM4e ± 1. The missing VO bands longwards of 8521Å put a strong upper limit on the spectral type to be earlier than dM5.5e.

Figure 5 shows prominent spectral features of G124-62A in comparison to AD Leo, a well-known dM3.5e [4] flare star. Although at first glance both spectra appear nearly identical, all metal lines are stronger in G124-62A (higher metallicity ?), while the emission features are slightly weaker. The TiO bands at 7087Å and 7125Å are stronger, implying a later spectral type than dM3.5e. Therefore we finally conclude that the spectral type is dM4.5e ± 0.5 subclasses. Since G124-62A is not known for showing variability in the optical and its x-ray flux is quite low, it is unlikely that this object is a flare star.

---

[4] see (Cincunegui & Mauas, 2004) for most recent spectral type determination

### 3.2. Age determination

In this section we derive the age of G124-62A by using different methods.

(1) The radial velocity of G124-62A deduced from our spectrum is $RV = -29.3 \pm 0.14$ km/s using the cross-correlation technique[5]. Together with the previously determined proper motion and the distance of the system, we calculated the UVW space velocities to $U = 35.33 \ldots 47.86$ km/s, $V = -23.28 \ldots -7.99$ km/s and $W = -10.85 \ldots -1.09$ km/s. This agrees well with the space motion of the Hyades supercluster of $U = +41.4 \pm 0.4$ km/s, $V = -18.5 \pm 0.9$ km/s and $W = -1.9 \pm 1.1$ km/s (Eggen, 1984). Montes et al. (2001) gave two criteria (compiled from Eggen's original criteria) that have to be fulfilled for a member of a supercluster:

(a) The inverse ratio of the components of the absolute proper motion in the direction of the convergent point ($\nu$) and perpendicular to it ($\tau$) should be smaller than 10%, weighted by the angular distance between the member candidate and the convergent point ($\lambda$) resulting in

$$\tau/\nu < 0.1 \sin^{-1}(\lambda). \qquad (3)$$

This means for G124-62 that more than 88% of the proper motion has to be directed towards the convergent point of the supercluster. The latter is given by (A,D)=($6^h\!.4, 6^\circ\!.5$) (Montes et al., 2001).

(b) The observed radial velocity should not differ by more than 4-8 km/s from a predicted radial velocity $\rho_c = V_T \cos \lambda$, where $V_T = 43.5$ km/s is the adopted supercluster velocity in the distance of G124-62.

Both criteria are fulfilled. The position angle between G124-62 and the convergent point is $\theta = 272.1$ deg, which differs from the direction of proper motion derived in Sect. 2.1 by only 1.8%. The observed radial velocity of $RV = -29.3 \pm 0.14$ km/s is larger than the predicted value of $V_T \cos \lambda = -24.74$ km/s but within the defined limits of 4-8 km/s. Adopting G124-62 as a member of the Hyades supercluster allows a first estimate for the age of the system given by the age of the cluster. Eggen (1992) gives an age between 300 and 800 Myr, even though older cluster members are known.

(2) The rotational velocity of G124-62A can be determined from our spectrum to $v \sin i = 12 \pm 1.5$ km/s. This value is significantly higher than the typical rotational velocity for single (old) dMe stars in the solar neighbourhood (Pettersen, 1991; Marcy & Chen, 1992), especially when the assumption holds that G124-62A is not a flare star. On the other hand, the measured $v \sin i$ lies well within the $v \sin i$ distribution of M-stars in the Hyades (625 Myr; Perryman et al. 1998), which shows an upper limit at 30 km/s, while the same distribution for the Pleiades (125 Myr; Stauffer, Schultz, & Kirkpatrick, 1998) includes M-stars with $v\sin i$ values of well over 80 km/s (Terndrup et al., 2000). Hence it is likely that G124-62 is older than 125 Myr but not much older than 625 Myr.

---

[5] Bailer-Jones (2004) measures a radial velocity of $-27.9 \pm 1.2$ km/s for G124-62B, which is in good agreement with our measured value for G124-62A, adding another proof for the physical connection of both objects.

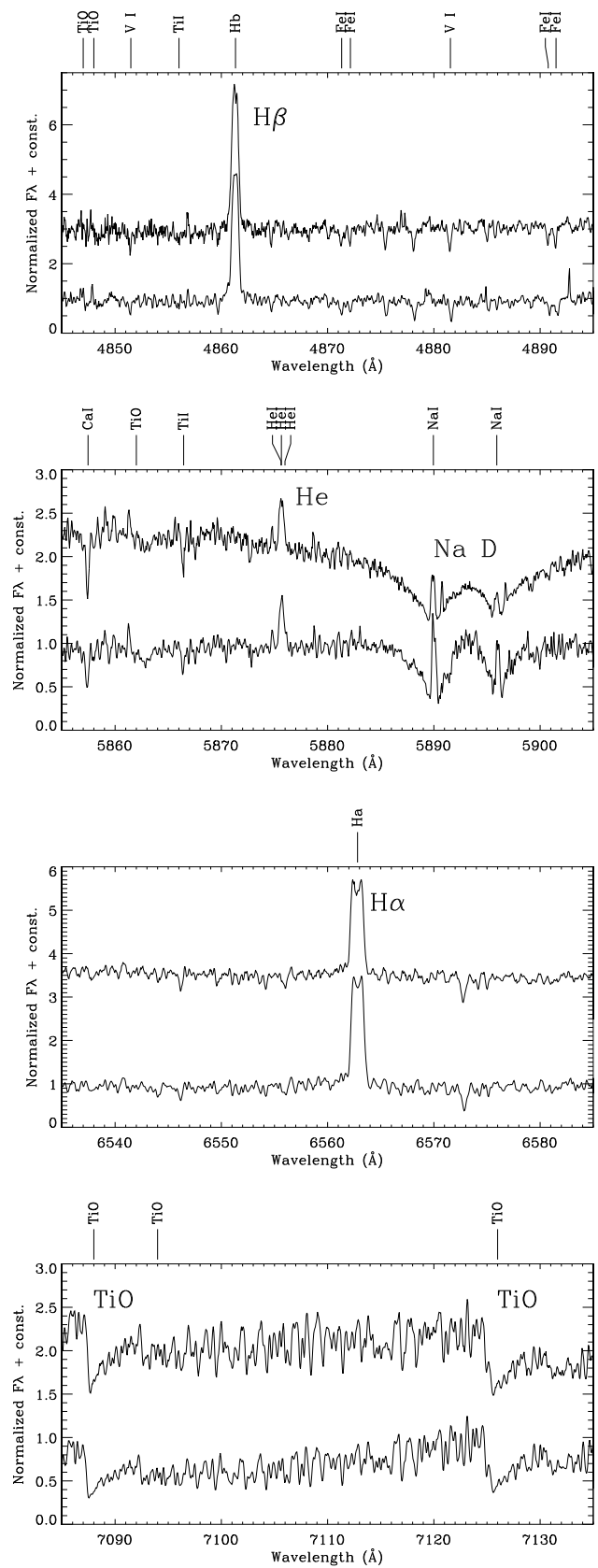

**Fig. 5.** Prominent features in the spectra of AD Leo (*lower spectrum*) and G124-62A (*upper spectrum*). See text for discussion.

(3) From the absence of a source of x-ray emission on the position of G124-62A in the ROSAT all-sky survey we can conclude an upper limit in x-ray flux of $5 \times 10^{-13}$ erg cm$^{-2}$ s$^{-1}$ for our object, resulting in an upper limit for the x-ray luminosity of $\log L_X = 28.84^{+0.16}_{-0.20}$ erg s$^{-1}$ at a distance of $34 \pm 7$ pc. Stelzer & Neuhäuser (2001) show that only $\sim 45\%$ of all stars in the Pleiades have an x-ray luminosity of $\leq 28.84$, but $\sim 85\%$ of the same population in the Hyades have such a low luminosity. We can again conclude that G124-62A seems older than the Pleiades (125 Myr) but is less likely to be older than the Hyades (625 Myr).

While the approaches so far have been based on direct age-activity relationships that show significant scatter and uncertainty, Hawley, Reid, & Tourtellot (2000) examined M dwarfs in open clusters and found a well-defined, age-dependent $V-I_C$ colour at which activity becomes dominant. This results in an upper age limit for a dMe star of a certain $V-I_C$ colour. Gizis, Reid, & Hawley (2002) give a relationship of

$$V - I_C = -6.91 + 1.05 \times \log(\text{age [Myr]}). \qquad (4)$$

For G124-62a we found $V$ and $I$ magnitudes of $m_V = 12.5 \pm 0.1$ mag from ASAS light-curve measurements and $m_I = 10.16 \pm 0.02$ mag from a preliminary version of the DENIS catalogue. We set $I_C = I_{DENIS}$, since the DENIS photometric system and the standard Cousins-CIT system differ by less than 0.05 mag for the $I$ band (Phan-Bao et al., 2003, see references therein). This results in a $V - I_C$ colour of $2.34 \pm 0.12$ mag, thus an upper age limit of $\sim 645^{+200}_{-150}$ Myr, which is consistent with the previously determined age values for G124-62A and the lower age limit of 0.5 Gyr for G124-62B, as concluded from the constraints for the age due to the absence of lithium in the spectrum of G124-62B (Martín et al., 1999). We therefore finally conclude that the age of the system is between 500 and 850 Myr and that G124-62 is a member of the Hyades supercluster.

## 4. Masses of the companions

To conclude on the mass of the companions, we use the DUSTY models from Chabrier, Baraffe, Allard, & Hauschildt (2000), since they reproduce the observed $(J - K)$ colour better than other models. We derive the luminosity of G124-62Ba and Bb from the previously determined absolute $K$ band magnitude and a bolometric correction $BC_K$ given by Leggett et al. (2001) to be

$$\begin{aligned} BC_K = &-0.31 + 5.124(J-K) - 2.03(J-K)^2 \\ &+ 0.13877(J-K)^3 \,, \ \sigma = 0.05 \end{aligned} \qquad (5)$$

for $0.75 \leq (J-K) \leq 1.6$, where $J$ and $K$ are in the UKIRT photometric system. We correct for the difference between the 2MASS and the UKIRT photometric system using the colour transformations given in Carpenter (2001):

$$\begin{aligned} (J-K)_{UKIRT} = &(0.935 \pm 0.010)(J-K_s)_{2MASS} \\ &+ (0.011 \pm 0.006) \end{aligned} \qquad (6)$$

$$\begin{aligned} K_{UKIRT} = &(K_s)_{2MASS} - (0.004 \pm 0.006)(J-K)_{UKIRT} \\ &- (0.002 \pm 0.004) \end{aligned} \qquad (7)$$

From $(J-K_s)_{2MASS} = 1.36 \pm 0.06$ mag and $(K_s)_{2MASS} = 10.77 \pm 0.48$ mag we derive $(J-K)_{UKIRT} = 1.28 \pm 0.08$ mag and $K_{UKIRT} = 10.77 \pm 0.49$ mag, implying $BC_K = 3.21 \pm 0.10$ mag for the bolometric correction. From these values we derive the luminosity of each component of G124-62B to be

$$\log(L/L_\odot) = -3.70 \pm 0.24 \,. \qquad (8)$$

Figure 6 shows the luminosity for various low mass stars and brown dwarfs as a function of age as predicted from the DUSTY models. We plotted age and luminosity for G124-62Ba and Bb with their uncertainties as a box that cover tracks from $\sim 0.054$ M$_\odot$ to $\sim 0.082$ M$_\odot$. Hence, both objects have masses near or on the hydrogen burning minimum mass (HBMM), which is $M_{HBMM} \simeq 0.070$ M$_\odot$ for the DUSTY model.

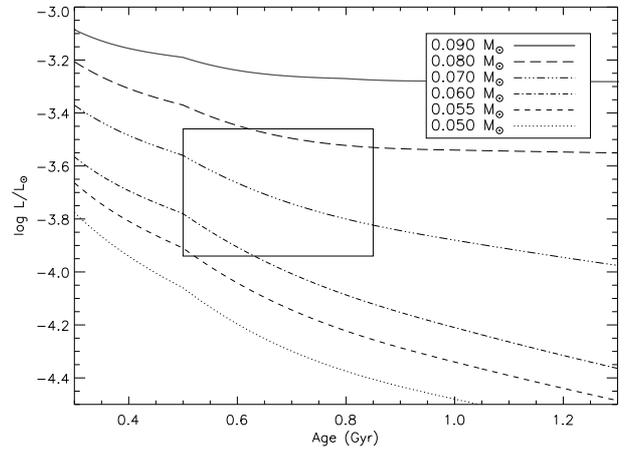

**Fig. 6.** Luminosity as a function of age for various low mass objects and brown dwarfs from DUSTY models. G124-62Ba and Bb are marked as a box.

## 5. Conclusions

We have demonstrated that G124-62 and DENIS-P J1441-0945 form a common proper motion pair, referred to as G124-62 A and G124-62B. Bouy et al. (2003) showed that the latter is a double, so we refer to them as G124-62Ba and Bb. From spectrophotometric measurements we determined a distance of the system of $d = 34 \pm 7$ pc, which is consistent with the measured parallax of $\pi = 30^{+17}_{-13}$ mas ($d = 33^{+24}_{-12}$ pc). A high resolution spectrum of G124-62A revealed a spectral type of dM4.5e $\pm$ 0.5. From x-ray luminosity, rotational velocity, $V - I_C$ colour-age relationship, and from the absence of lithium in the spectrum of G124-62B, we determined the age of the system to 500

.. 850 Myr. We could finally derive the mass of G124-62Ba and Bb to $0.072^{+0.010}_{-0.018}$ $M_\odot$ for each component. Hence the companion could either be a very low mass star or a high-mass brown dwarf. A more precise parallax measurement is needed for better determination of distance and mass. We also find that G124-62 is a member of the Hyades supercluster. It is thus interesting to note that this is the third system in the Hyades known of where an M star is orbited by one or two very low mass stars or brown dwarfs (Guenther et al., 2004).

*Acknowledgements.* We would like to thank Laurent Cambresy, Observatoire de Strasbourg, for the preliminary photometric data on G124-62A from the Deep Near Infrared Survey of the Southern Sky (DENIS) and Brigitte König, MPE Garching, for her helpful comments on flare stars. Finally we thank the referee, Gillian Knapp, for her helpful comments on the paper.

This publication made use of data products from the Two Micron All Sky Survey (which is a joint project of the University of Massachusetts and the Infrared Processing and Analysis Center/California Institute of Technology, funded by the National Aeronautics and Space Administration and the National Science Foundation) and of data based on the SuperCOSMOS Sky Surveys at the Wide Field Astronomy Unit of the Institute of Astronomy, University of Edinburgh. We also used the VizieR catalogue access tool, CDS, Strasbourg.